\pdfoutput=1

\documentclass[onecolumn,superscriptaddress]{revtex4}

\usepackage{graphicx,float}
\usepackage{epsfig}
\usepackage{epstopdf}

\usepackage{amsmath}
\usepackage{amssymb}
\usepackage{bm}
\usepackage{color}
\usepackage[usenames,dvipsnames]{xcolor}
\usepackage[colorlinks=true,citecolor=blue,urlcolor=blue]{hyperref}

\usepackage{hyperref}
\hypersetup{colorlinks=true,linkcolor=blue,citecolor=blue,urlcolor=blue}

\newcommand{\be}{\begin{equation}}
\newcommand{\ee}{\end{equation}}

\newcommand{\sket}[1]{{\ensuremath{\lvert#1\rangle}}}
\newcommand{\lket}[1]{{\ensuremath{\left\lvert#1\right\rangle}}}
\newcommand{\ket}[1]{\if@display\lket{#1}\else\sket{#1}\fi}

\newcommand{\sbra}[1]{{\ensuremath{\langle#1\rvert}}}
\newcommand{\lbra}[1]{{\ensuremath{\left\langle#1\right\rvert}}}
\newcommand{\bra}[1]{\if@display\lbra{#1}\else\sbra{#1}\fi}

\newcommand{\sbraket}[2]{{\ensuremath{\langle#1\rvert#2\rangle}}}
\newcommand{\lbraket}[2]{{\ensuremath{\left\langle#1\!\left\rvert\vphantom{#1}#2\right.\!\right\rangle}}}
\newcommand{\braket}[2]{\if@display\lbraket{#1}{#2}\else\sbraket{#1}{#2}\fi}

\newcommand{\sketbra}[2]{{\ensuremath{\lvert #1\rangle\!\langle #2\rvert}}}
\newcommand{\lketbra}[2]{{\ensuremath{\left\lvert #1\right\rangle\!\!\left\langle #2\right\rvert}}}
\newcommand{\ketbra}[2]{\if@display\lketbra{#1}{#2}\else\sketbra{#1}{#2}\fi}

\begin{document}

%%%%%%%%%%%%%%%%%%%%%%%%%%%%%%%%%%%%%%%%%%%%%%%%%%%%%%%%%%%%%%%%%%%

\title{Quantum random number generation enhanced by weak-coherent states interference}

%%%%%%%%%%%%%%%%%%%%%%%%%%%%%%%%%%%%%%%%%%%%%%%%%%%%%%%%%%%%%%%%%%%

\date{\today}

\author{T. Ferreira da Silva}
\affiliation{Optical Metrology Division, National Institute of Metrology, Quality and Technology, 25250-020 Duque de Caxias RJ, Brazil}
\email{tfsilva@inmetro.gov.br}

\author{G. B. Xavier}
\affiliation{Departamento de Ingenier\'{i}a El\'{e}ctrica, Millenium Nucleus of Advanced Optics and Center for Optics and Photonics, Universidad de Concepci\'{o}n, Casilla 160-C Concepci\'{o}n, Chile}

\author{G. C. Amaral}
\affiliation{Center for Telecommunications Studies, Pontifical Catholic University of Rio de Janeiro, 22451-900 Rio de Janeiro RJ, Brazil}

\author{G. P. Tempor\~{a}o}
\affiliation{Center for Telecommunications Studies, Pontifical Catholic University of Rio de Janeiro, 22451-900 Rio de Janeiro RJ, Brazil}

\author{J. P. von der Weid}
\affiliation{Center for Telecommunications Studies, Pontifical Catholic University of Rio de Janeiro, 22451-900 Rio de Janeiro RJ, Brazil}

%%%%%%%%%%%%%%%%%%%%%%%%%%%%%%%%%%%%%%%%%%%%%%%%%%%%%%%%%%%%%%%%%%%

\begin{abstract}
We propose and demonstrate a technique for quantum random number generation based on the random population of the output spatial modes of a beam splitter when both inputs are simultaneously fed with indistinguishable weak coherent states. We simulate and experimentally validate the probability of generation of random bits as a function of the average photon number per input, and compare it to the traditional approach of a single weak coherent state transmitted through a beam-splitter, showing an improvement of up to 32\%. The ensuing interference phenomenon reduces the probability of coincident counts between the detectors associated with bits 0 and 1, thus increasing the probability of occurrence of a valid output. A long bit string is assessed by a standard randomness test suite with good confidence. Our proposal can be easily implemented and opens attractive performance gains without a significant trade-off.

\end{abstract}

%%%%%%%%%%%%%%%%%%%%%%%%%%%%%%%%%%%%%%%%%%%%%%%%%%%%%%%%%%%%%%%%%%%

%%%%%%%%%%%%%%%%%%%%%%%%%%%%%%%%%%%%%%%%%%%%%%%%%%%%%%%%%%%%%%%%%%%

\maketitle

%%%%%%%%%%%%%%%%%%%%%%%%%%%%%%%%%%%%%%%%%%%%%%%%%%%%%%%%%%%%%%%%%%%

\section{Introduction}
Random numbers are an important resource for several applications in science and engineering, as in the Monte-Carlo modeling method \cite{MetropolisJASA49} and cryptography \cite{GisinRMP02}, and also for daily applications, as lotteries and gambling. Random number generators (RNGs) based on a deterministic process, usually an algorithm, generates a sequence that appears to be random at a first glance -- in the sense of uniformly distributed statistics -- but, in spite of being successfully employed in simulation tools \cite{site:mathworks}, each generated bit is predictable in principle. This is a drawback that makes this kind of RNG inappropriate for some sensitive applications, as in quantum key distribution (QKD). Quantum random number generators (QRNGs), on the other hand, are based on intrinsically random quantum mechanical processes and, hence, are considered truly random devices \cite{HerreroArxiv16}. 

A very elegant implementation for a QRNG is the path-splitting of photons in a beam splitter (BS) \cite{RarityJMO94,StefanovJMO00,JenneweinRSI00} and in fact constitutes the operational principle of popular commercially available devices \cite{HerreroArxiv16}. A single photon impinging on a symmetrical BS has identical probabilities of emerging at either of its output modes -- the state is thus in a superposition of both spatial modes.  If one places one detector at each output of the BS, bits ``0'' and ``1'' can be assigned to a valid detection on either side. Faint optical sources are usual in these generators due to their simplicity and efficacy, but the emission of multiple photons limits the maximum probability of generating a random bit because it leads to collisions -- coincidence events between the detectors -- that are neither assigned as ``0'' nor ``1'' and thus discarded. An optimum operation point is thus obtained from the trade-off of avoiding multi-photon states while reducing the vacuum component.

Photon sources based on spontaneous parametric down-conversion could be used to reduce the collision probability by avoiding multi-photon emission in a sub-Poisson emission regime \cite{MaCPL04} or by producing path entangled photons \cite{KwonAO09}. This kind of source, however, has low brightness and is significantly more complex to implement than faint sources. 

In this paper we present a technique for enhancing a QRNG based on randomly populating the output spatial modes of a beam splitter by feeding it with indistinguishable mutually-incoherent weak coherent states (WCSs). The interference phenomenon (\cite{OuBOOK,HongPRL86,PaulRMP86,KimPRA13,ThiagoJOSAB15,GustavoOL16}, and references therein) plays the role of reducing the probability of occurrence of a collision. This simple modification from the traditional splitting of a single WCS enhances the probability of generating a valid random bit. We experimentally validate the theoretical model and successfully assess the generated bit sequence using standard randomness tests, proving our proposal to be an attractive approach for a practical QRNG.

\section{Path-splitting QRNG enhanced by two-photon interference}
In a standard path--splitting QRNG, photons are sent into a symmetric beam splitter (BS$_1$) and a valid bit is assigned whenever only one of the two single-photon detectors (SPDs) -- each placed at one of the output modes -- clicks. When dealing with states with more than one photon at the input of the BS -- as in the case of WCSs weak coherent states -- the coincident counts between the detectors limit the maximum generation probability of the QRNG, as collision events are discarded. Feeding the BS with coherent state and vacuum at the input modes $a$ and $b$ results in coherent states at the output modes $c$ and $d$: $\vert\alpha,0\rangle_{a,b} \rightarrow \left\vert\frac{j\alpha}{\sqrt{2}},\frac{\alpha}{\sqrt{2}}\right\rangle_{c,d}$; each photon entering the beam splitter takes a random, independent output. The probability of simultaneously finding $m$ and $n$ photons at spatial modes $c$ and $d$ thus follows the Poisson distribution \cite{OuBOOK}. 

\begin{figure}[htbp]
\centering
\includegraphics[width=0.6\textwidth]{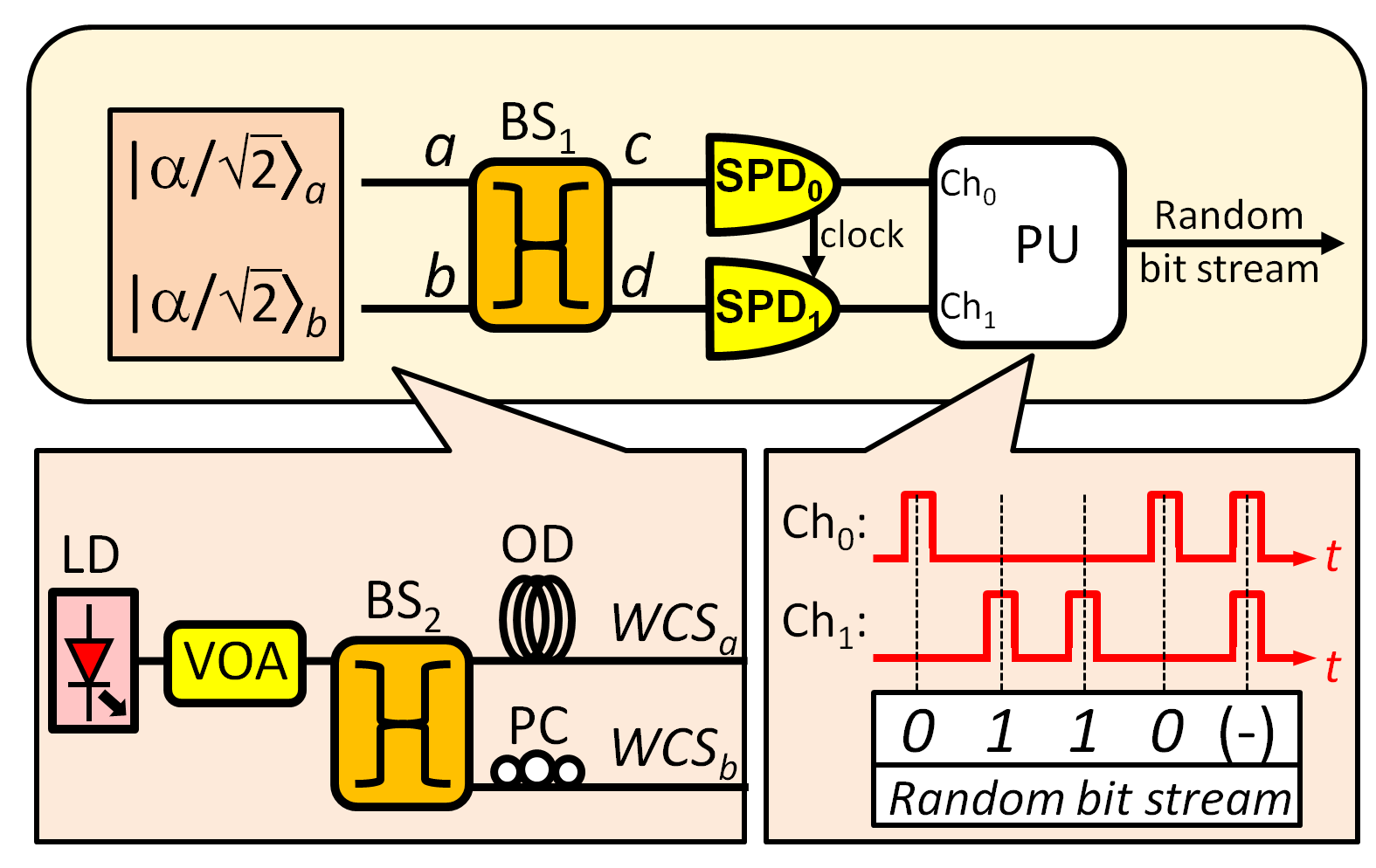}
\caption{QRNG based on random path-splitting enhanced by interference of indistinguishable WCSs. The experimental setup of the optical source is detailed in the inset on the left. The inset on the right shows the sampled channels: (-) represents no output bit generated when both SPDs click, while a single-channel detection is associated to bit 0 or 1. LD: laser diode; VOA: variable optical attenuator; BS: beam splitter; OD: optical delay; PC: polarization controller; PU: processing unit.}
\label{fig:1}
\end{figure}

Our proposal is depicted in Fig. \ref{fig:1}. Two-photon interference takes a major role in our proposed scheme since the source of randomness is not only the indivisibility of the light quanta, but the bosonic nature of the photons: two indistinguishable photons that impinge on a BS emerge bunched together in a \textit{random} output mode \cite{OuBOOK}. This means the output state is in a superposition of both modes: $\vert 1,1\rangle_{a,b}\rightarrow\left(\vert 2,0\rangle_{c,d}+\vert 0,2\rangle_{c,d}\right)/\sqrt{2}$. 

Two indistinguishable WCSs $\vert \sqrt{\mu}e^{j\theta}\rangle_{a,b}$ at the input modes of the beam splitter, with uniformly distributed random phases $\theta$ and equal amplitudes $\sqrt{\mu}$, are described by the density operator

\begin{equation}
\rho_{a,b} = \left[\int_0^{2\pi} \left\vert\sqrt{\mu}e^{j\theta}\right\rangle\left\langle\sqrt{\mu}e^{j\theta}\right\vert d\theta \right]^{\otimes 2}
\label{eq:1}
\end{equation}
where $\mu$ is the average number of photons per time interval. By indistinguishable states we mean the polarization, spatial and temporal modes are matched and they have identical values of $\mu$. This state can be decomposed into pairs of Fock states $\vert m,n\rangle_{a,b}$, where the probability of occurring $m$ photons in mode $a$ and $n$ photons in mode $b$ follows the Poisson distribution

\begin{equation}
P_{a,b}\left(m,n\right)=e^{-\mu}\mu^{m+n}/\left(m!n!2^{m,n}\right)
\label{eq:2}
\end{equation}
Each input state $\vert m,n\rangle_{a,b}$ results in the output given by \cite{OuBOOK}

\begin{equation}
\vert\psi\rangle_{c,d}= \sum\limits_{u=0}^m\sum\limits_{v=0}^n
\frac{j^{u+v}\sqrt{m!n!\left(m-u+v\right)!\left(n-v+u\right)!}}{2^{\left(m+n\right)/2}\left(m-u\right)!u!\left(n-v\right)!v!} \left\vert m-u+v,n-v+u\right\rangle_{c,d}
\label{eq:3}
\end{equation}

The probability of simultaneously finding $M$ photons at mode $c$ and $N$ photons at mode $d$ is obtained by summing over the projections of the state in eq. (\ref{eq:3}) weighted by the probability of each input to occur:

\begin{equation}
P_{c,d}(M,N)=\sum_{m=0}^{\infty}\sum_{n=0}^{\infty}P_{a,b}(m,n)\langle\psi\vert M,N\rangle_{c,d}
\label{eq:4}
\end{equation}

The maximum contrast of the coincidence counts between modes $c$ and $d$ is limited to 0.5 in the case of indistinguishable WCSs as inputs due to the finite probability of occurring multi-photon events at the inputs \cite{ThiagoJOSAB15,GustavoOL16}.

For perfect single-photon detectors, any non-vacuum number state reaching the device originates a detection event. In practice, however, there is a finite probability of detection of photons at modes $c$ and $d$: an $i$-photon state is detected by a (threshold) SPD of overall efficiency $\eta$ with probability $\eta_i=1-\left(1-\eta\right)^i$ \cite{MaPRA05}. This is applied to each output mode, according to the number of photons. The probability of generating a valid bit, $P_{gen}$, is therefore computed from the events with any number of photons being counted in only one mode. The probability of an event being discarded due to the simultaneous detection at both output modes, $P_{disc}$, represents a collision. This is computed, neglecting the SPDs dark counts, as

\begin{equation}
\begin{array}{rl}
P_{gen}=&\sum\limits_{M=1}^\infty P_{c,d}\left(M,0\right)\eta_M+
\sum\limits_{M=1}^\infty\sum\limits_{N=1}^\infty P_{c,d}\left(M,N\right)\eta_M\left(1-\eta_N\right)+\\
        &\sum\limits_{N=1}^\infty P_{c,d}\left(0,N\right)\eta_N+
\sum\limits_{M=1}^\infty\sum\limits_{N=1}^\infty P_{c,d}\left(M,N\right)\left(1-\eta_M\right)\eta_N \\
P_{disc}=&\sum\limits_{M=1}^\infty\sum\limits_{N=1}^\infty P_{c,d}\left(M,N\right) \eta_M\eta_N
\end{array}
\label{eq:5}
\end{equation}

The throughput of the generator using continuous-wave light is obtained by multiplying the probability of generating a valid bit by the detection gate frequency.

\section{Experimental setup}
The continuous-wave emission of a telecom DFB laser source at 1547.8 nm is split into two branches by an optical beam splitter ($BS_2$), as shown in the detail of Fig. \ref{fig:1}. The signals are decorrelated from each other by an 8-km long optical fiber spool (this spool was chosen to be much longer than the mutual coherence length of the sources, measured as 5.2 m) and sent to BS$_1$ as two mutually-incoherent independent sources. The two states are power balanced -- resulting in identical values of $\mu$ -- and their polarization modes, overlapped. The average number of photons per time interval of both states is simultaneously adjusted by the variable optical attenuator (VOA) placed right after the laser source. The all-fiber setup guarantees the spatial modes overlap. 

The output modes of BS$_2$ are connected to an InGaAs/InP avalanche photodiode-based SPD. The detectors operate in gated Geiger mode and are synchronously triggered by a periodic 100 kHz clock signal. At each trigger pulse, the detectors open a 1.6-ns wide temporal gate window with 15\% detection efficiency. Afterpulses are avoided by the low gate frequency \cite{ThiagoJQE11}, but an aftercount deadtime can be employed if higher rate is desired with this kind of detector. Maximum interference between the WCSs is ensured with matched states and verified at low $\mu$ by scanning the relative delay between the detection gates of the SPDs, resulting in about 49\% visibility.

The detection pulses generated by detectors SPD$_0$ and SPD$_1$ are sent to the processing unit (PU), composed by a two-channel 100 MSamples/s acquisition board driven by a personal computer. Both channels are simultaneously sampled, fetched and processed in blocks. If a detection pulse is sampled at only one channel, this is treated as a valid bit, 0 and 1 associated to each channel. Whenever an event occurs at both channels simultaneously, it is considered a collision and is discarded. A predefined temporal matching window is used to account for an eventual offset delay between both channels. The QRNG runs until the desired number of bits is obtained.

The attenuation set at the VOA is varied and the statistics of events is obtained using indistinguishable WCSs. The experiment is repeated with a single WCS by connecting the faint laser source to one input port of the main beam splitter with vacuum at the other.

\section{Results}
The probability of our proposed QRNG to generate a valid output bit or to discard the collision event was numerically simulated.  For practical purposes, the probability to generate or discard of bits when using a detector with overall detection efficiency $\eta <$1 can be computed by replacing $\mu$ for $\mu\eta$. Different values of $\mu\eta$ at the input modes of the beam splitter (BS$_1$ at Fig. $\ref{fig:1}$) were employed using eq. (\ref{eq:4}). The case of a single WCS plus vacuum is used as a benchmark for evaluating the improvement offered by our proposal on the bit generation probability. The probability of simultaneously finding $M$ and $N$ photons at the output of the BS in this case is directly obtained from the Poisson distribution \cite{OuBOOK} $P_{c,d}\left(M,N\right)=e^{-\mu}\mu^{M+N}/\left(M!N!2^{M+N}\right)$. The maximum number of photons used for the expansion of the WCSs keeps the cumulative distribution function of the Poisson distribution greater than 99.9\%. 

\begin{figure}[htbp]
\centering
\includegraphics[width=0.6\textwidth]{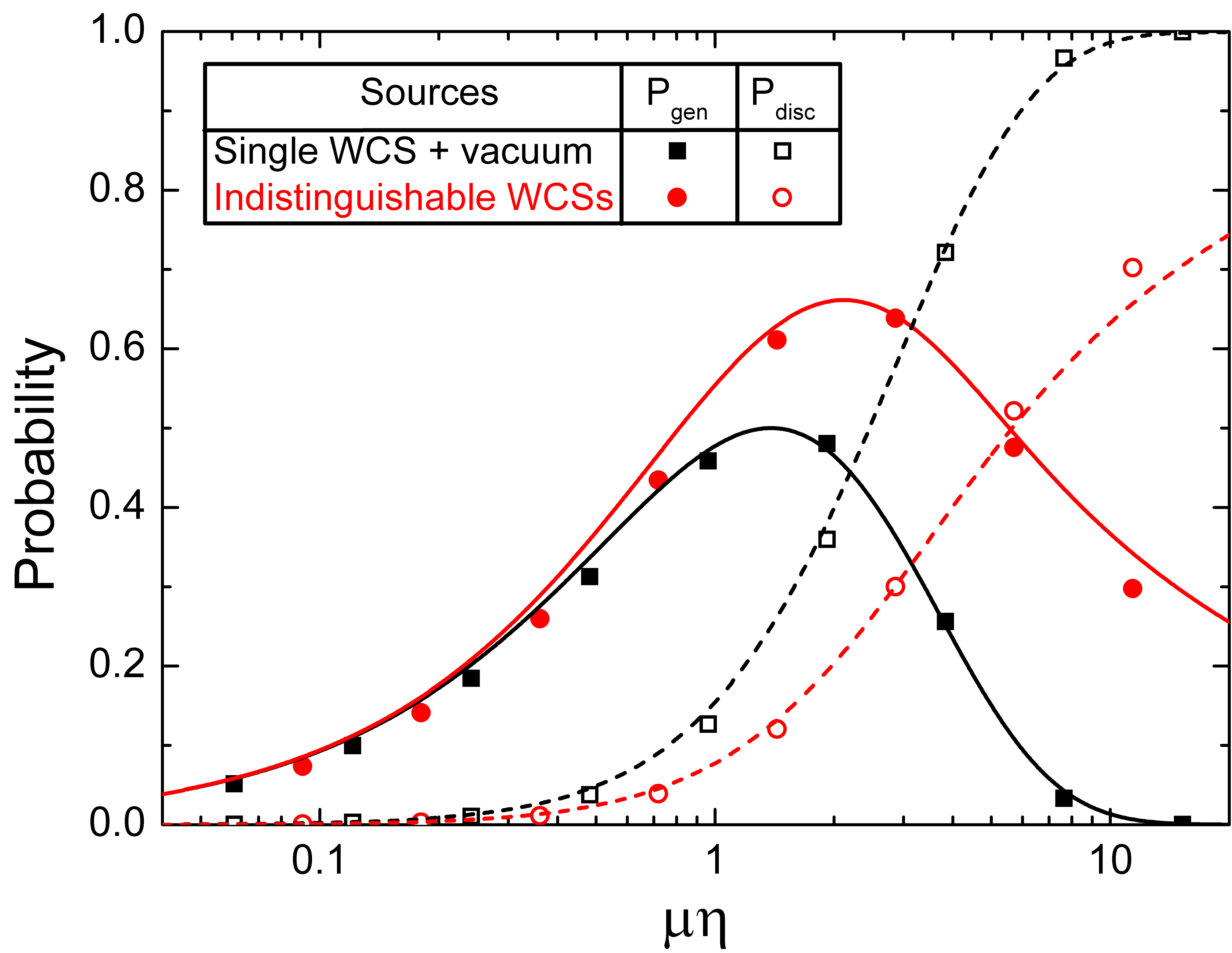}
\caption{Probability of generation of a valid output bit ($P_{gen}$) and probability of discard due to collision ($P_{disc}$) by inputting single WCS plus vacuum or indistinguishable WCSs into the BS. Points are measured data and lines are obtained by numerical simulation.}
\label{fig:2}
\end{figure}

The results, shown in Fig. \ref{fig:2}, reveal that the maximum probability of generating a valid random bit per detection gate can be enhanced when two indistinguishable WCSs are used. The experimental results, also presented in Fig. \ref{fig:2}, clearly agree with the simulation, validating the model. In the experiment, the average number of photons at the input of the interferometer was varied to have a $\mu\eta$ product in the range from 0.05 up to about 20. The number of events occurring at only one channel and also events simultaneously taken at both detectors were counted. The intrinsic decay in the visibility of the interference with the increase of $\mu\eta$, as the multi-photon states become more probable, is accounted for by the model and naturally embedded in the experimental results. Degradation of the interference caused by misalignment of the WCSs -- see \cite{GustavoOL16} -- would reduce the gain of our QRNG over the path-splitting-only method, which tends to zero with fully-distinguishable states.

The highest probability of generating a valid output is 0.50 at $\mu\eta\approx$1.4 for the single WCS plus vacuum, and 0.66 at $\mu\eta\approx$2.1 when using indistinguishable WCSs -- $\mu$ considered as the total number of photons at the input of the beam splitter. The improvement ratio, considering both cases at their optimum condition, is of a factor of 1.32. Even if non-ideal detectors are considered, with $\eta < 1$, the maximum generation probability can be achieved in both cases by increasing the average number of photons per time interval, meaning that the QRNG performance is not affected by the detection efficiencies or internal optical losses.

Figure \ref{fig:2} also shows the saturation of the coincident counts between both SPDs with the increase in $\mu\eta$, up to the trigger rate. For high values of $\mu\eta$, both channels register events with high probability and the number of coincidences naturally increases. With indistinguishable WCSs this increase builds up at at higher values of $\mu\eta$ due to the two-photon interference effect. Even though it eventually saturates, the maximum generation probability of a valid event is higher than that when using a single WCS plus vacuum.

\section{Randomness tests}
A long random sequence was experimentally generated using our proposed approach of indistinguishable WCSs with the generator set to the maximum probability of generating a valid output. The path-splitting generator is naturally susceptible of producing a biased output, as the transmittance after the BS and the detection efficiency of the SPDs are not perfectly balanced. To equalize the probabilities of generating a zero or a one , originally measured with as 0.49 and 0.51 respectively, we used the von Neumann approach \cite{PeresAS92}. However any de-biasing method used with path-splitting RNGs, including more efficient ones \cite{PeresAS92}, can also be used here.

\begin{figure}[htbp]
\centering
\includegraphics[width=0.6\textwidth]{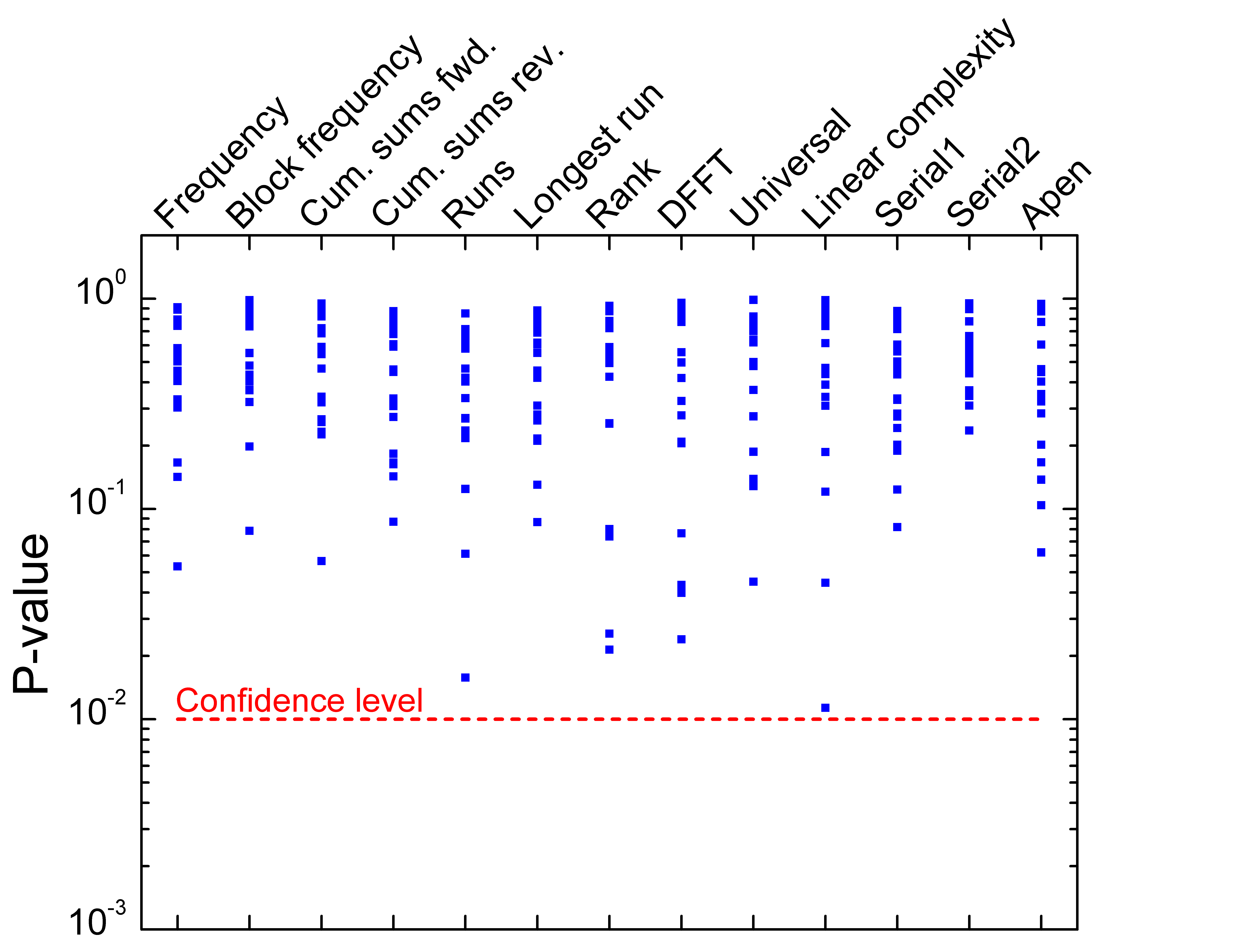}
\caption{Results of the randomness test suite for the generated random bit sequence using indistinguishable WCSs.}
\label{fig:3}
\end{figure}

The produced sequence was tested after ubiasing with the standard NIST randomness test suite (version 1.8) \cite{nist}. The 19-Mbits long sequence was split into 19 blocks of $10^6$ bits and the results of the tests are reported in Fig. \ref{fig:3} as p-values, which indicate randomness if above the confidence level of 0.01.

\section{Conclusion}

We report a quantum random number generation method that, using interference of indistinguishable mutually-incoherent weak coherent states, enhances the bit generation probability by up to 32\% when compared to the traditional splitting of a single weak coherent state. Whereas this last approach is solely based on the indivisibility of the light quantum, ours also benefits from the bosonic nature of the photons to reduce the coincident events between the output modes (that are discarded). A long bit stream is generated and fed to a standard random number test suite, with the results indicating good confidence in the randomness of the string. Our proposal proves to be a practical, attractive and simple approach to improve the generation of random bits with the inclusion of an extra beam splitter and an optical delay to the traditional approach of a single coherent state (plus vacuum) impinging on a beam splitter.

{\it Acknowledgements.---}G. B. X. acknowledges support of FONDECYT no. 1150101, CONICYT PFB08-024 and Milenio RC130001. G. C. A., G. P. T. and J. P. W. acknowledge support of CPNq.

%%%%%%%%%%%%%%%%%%%%%%%%%%%%%%%%%%%%%%%%%%%%%%%%%%%%%%%%%%%%%%%%%%%

%%%%%%%%%%%%%%%%%%%%%%%%%%%%%%%%%%%%%%%%%%%%%%%%%%%%%%%%%%%%%%%%%%%


\begin{thebibliography}{000}

\bibitem{MetropolisJASA49}
N. Metropolis and S. Ulam,
``The Monte Carlo method,''
J. Am. Stat. Assoc. \textbf{44}, 335--341 (1949).

\bibitem{GisinRMP02}
N. Gisin, G. Ribordy, W. Tittel, and H. Zbinden,
``Quantum cryptography,''
Rev. Mod. Phys. \textbf{74}, 145--195 (2002).

\bibitem{site:mathworks}
http://www.mathworks.com/products/matlab/

\bibitem{HerreroArxiv16}
M. Herrero-Collantes and J. C. Garcia-Escartin,
``Quantum random number generators,''
arXiv:1604.03304v1, 2016.

\bibitem{RarityJMO94}
J. Rarity, P. Owens, and P. Tapster,
``Quantum random-number generation and key sharing,''
J. Mod. Opt. \textbf{41}, 2435--2444 (1994).

\bibitem{StefanovJMO00}
A. Stefanov, N. Gisin, O. Guinnard, L. Guinnard, and H. Zbinden,
``Optical quantum random number generator,''
J. Modern Optics \textbf{47}, 595--598, 2000.

\bibitem{JenneweinRSI00}
T. Jennewein, U. Achleitner, G. Weihs, H. Weinfurter, and A. Zeilinger,
``A fast and compact quantum random number generator,''
Rev. Sci. Instrum. \textbf{71}, 1675–-1680 (2000).

\bibitem{MaCPL04}
H.-Q. Ma, S.-M. Wang, D. Zhang, J.-T. Chang, L.-L. Ji, Y.-X. Hou, and L.-A. Wu, 
``A random number generator based on quantum entangled photon pairs,''
Chin. Phys. Lett. 21, 1961 (2004).

\bibitem{KwonAO09}
O. Kwon, Y.-W. Cho, and Y.-H. Kim,
``Quantum random number generator using photon-number path entanglement,''
App. Opt. \textbf{48}, 1774--1778 (2009).

\bibitem{OuBOOK}
Z.-Y. J. Ou,
\textit{Multi-Photon Quantum Interference}
(Springer, 2007).

\bibitem{HongPRL86}
C. K. Hong, Z. Y. Ou, and L. Mandel, 
``Measurement of subpicosecond time intervals between two photons by interference,''
Phys. Rev. Lett. \textbf{59}, 2044--2046 (1987).

\bibitem{PaulRMP86}
H. Paul,
``Interference between independent photons,''
Rev. Mod. Phys. \textbf{58}, 209 (1986).

\bibitem{KimPRA13}
Y.-S. Kim, O. Slattery, P. S. Kuo, and X. Tang,
``Conditions for two-photon interference with coherent pulses,''
Phys. Rev. A \textbf{87}, 063843 (2013).

\bibitem{ThiagoJOSAB15}
T. Ferreira da Silva, G. C. do Amaral, D. Vitoreti, G. P. Tempor\~{a}o, and J. P. von der Weid,
``Spectral characterization of weak coherent state sources based on two-photon interference,''
J. Opt. Soc. Am. B \textbf{32}, 545--549 (2015).

\bibitem{GustavoOL16}
G. C. Amaral, T. Ferreira da Silva, G. P. Tempor\~{a}o, and J. P. von der Weid
``Few-photon heteorodyne spectroscopy,''
Opt. Lett. \textbf{41}, 1502--1505 (2016).

\bibitem{MaPRA05}
X. Ma, B. Qi, Z. Zhao, and H.-K. Lo,
``Practical decoy state for quantum key distribution,''
Phys. Rev. A \textbf{72}, 012326 (2005).

\bibitem{ThiagoJQE11}
T. Ferreira da Silva, G. B. Xavier, and J. P. von der Weid,
``Real-time characterization of gated-mode single-photon detectors,''
IEEE J. Quantum Electron. \textbf{47}, 1251--1256 (2011).

\bibitem{PeresAS92}
Y. Peres,
``Iterating von Neumann's procedure for extracting random bits,''
Ann. Stat. \textbf{20}, 590--597 (1992).

\bibitem{nist}
A. Rukhin, J. Soto, J. Nechvatal, M. Smid, E. Barker, S. Leigh, M. Levenson, M.  Vangel, D. Banks, A. Heckert, J. Dray, and S. Vo,
``A statistical test suite for random and pseudorandom number generators for cryptographic applications,''
NIST Special Publication 800-22.

\end{thebibliography}
\end{document}